\documentclass[reprint,showpacs,preprintnumbers,superscriptaddress,pra,aps]{revtex4-2}

\usepackage{amsmath}
\usepackage{amssymb}
\usepackage{graphicx}
\usepackage{dcolumn}
\usepackage{bm}
\usepackage{color}
\usepackage{lineno}
\usepackage[all]{nowidow}
\usepackage{siunitx}[=v2]

\newcommand{\be}{\begin{equation}}
	\newcommand{\ee}{\end{equation}}

\newcommand{\fig}[1]{Fig.~\ref{#1}}

\newcommand{\Fig}[1]{Figure~\ref{#1}}

\newcommand{\eq}[1]{Eq.~\eqref{#1}}

\newcommand{\Eq}[1]{Equation~\eqref{#1}}

\newcommand{\bea}{\begin{eqnarray}}
	\newcommand{\eea}{\end{eqnarray}}
 
\newcommand{\ba}{\begin{array}}
	\newcommand{\ea}{\end{array}}

\newcommand{\bl}{\begin{flalign}}
	\newcommand{\enl}{\end{flalign}}

\newcommand{\braketmatrix}[3]{\left \langle #1 \middle| #2 \middle| #3 \right \rangle}

\begin{document}
	\author{Tobias Weitz}
    \email[]{tobias.weitz@fau.de}
	\affiliation{Department of Physics, Friedrich-Alexander-Universität Erlangen-Nürnberg (FAU), Staudtstrasse 1, D-91058 Erlangen, Germany}
	\author{Christian Heide} 
	\affiliation{Department of Physics, Friedrich-Alexander-Universität Erlangen-Nürnberg (FAU), Staudtstrasse 1, D-91058 Erlangen, Germany}
	\affiliation{Stanford PULSE Institute, SLAC National Accelerator Laboratory, Menlo Park, California 94025, USA}
	\author{Peter Hommelhoff}
    \email[]{peter.hommelhoff@fau.de}
	\affiliation{Department of Physics, Friedrich-Alexander-Universität Erlangen-Nürnberg (FAU), Staudtstrasse 1, D-91058 Erlangen, Germany}	
	\date{\today}
	
	\title{Strong-Field Bloch Electron Interferometry \\for Band Structure Retrieval}
	
	\begin{abstract}
	When Bloch electrons in a solid are exposed to a strong optical field, they are coherently driven in their respective bands where they acquire a quantum phase as the imprint of the band shape. If an electron approaches an avoided crossing formed by two bands, it may be split by undergoing a Landau-Zener transition. We here employ subsequent Landau-Zener transitions to realize strong-field Bloch electron interferometry (SFBEI), allowing us to reveal band structure information. In particular, we measure the Fermi velocity (band slope) of graphene in the vicinity of the K points as \SI{1.07(4)}{\nano\m\per\femto\s}. We expect SFBEI for band structure retrieval to apply to a wide range of material systems and experimental conditions, making it suitable for studying transient changes in band structure with femtosecond temporal resolution at ambient conditions. 
	\end{abstract}
	
	\maketitle
	
	The band dispersion and topology of a solid govern its optical and electrical properties, ranging from light absorption to electrical conductivity and phase transitions. Mainly because of its peculiar band structure, the optical and electrical properties of the semimetal graphene, for example, are exceptional and herald a number of potential applications~\cite{CastroNeto2009,Andrei2020,Block2021,Boolakee2022}. Direct access to the Bloch electrons residing in the bands of interest allows imaging their native band structure. This principle has been widely applied in angle-resolved photoemission spectroscopy (ARPES)~\cite{Sprinkle2009,Johannsen2013,Gierz2013}. A drawback of this technique is that in the required photoemission process the quantum mechanical properties of electrons usually remain concealed. This way, precious information beyond their probability distribution, such as their sensitivity to band topology cannot be recovered.
	
	Other recent approaches for band structure retrieval are based on coherent, ultrafast strong-field electron dynamics. They rely on coherently steering the quantum-mechanical electron wavefunction across the bands. Various avenues emerged to study the band structure as well as the band-specific quantum nature of the driven electrons: Band structure tomography via harmonic sideband~\cite{Borsch2020} and high-harmonic spectroscopy~\cite{Vampa2015a,Mitra2023} has been shown. High-harmonic emission and time-resolved ARPES enabled studying the electronic coherence properties~\cite{Heide2022a,Ito2023}. Recently, band topology-sensitive high harmonic emission has caught particular attention for both the fundamental understanding of solids as well as their potential in lightwave electronics~\cite{Schmid2021,Bai2021,Heide2022,Mitra2023,Borsch2023}. 
	
\begin{figure*} 
		\begin{center}
			\includegraphics[width=\linewidth]{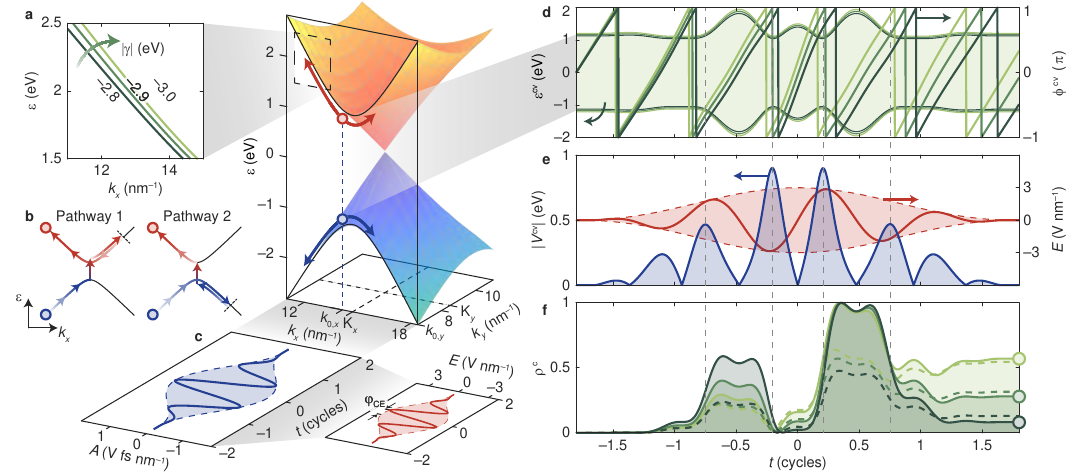}
            \caption{Field-driven electron dynamics in graphene and the influence of the hopping parameter $\gamma$. (a) Dirac cone-like dispersion around the K point with valence (blue) and conduction band (orange). The intraband trajectory of an exemplary electron-hole pair (red and blue circles) starting at $\mathbf{k}_0=[14.1,\,6.6]\,\si{\per\nano\m}$ and driven by the vector potential shown in (c) is shown as double arrows (blue in valence band, red in conduction band). The inset shows the influence of $\gamma$ on the band dispersion: With increasing $\left|\gamma\right|$, the linear slope, i.e., the Fermi velocity $v_\text{F}=\tfrac{3}{2}\left|\gamma\right|a$ increases. For instance, at $\gamma=\SI{-2.8}{\eV}$, $v_\text{F}=\SI{1.03}{\nano\m\per\femto\s}$ whereas at $\gamma=\SI{-3.0}{\eV}$, $v_\text{F}=\SI{1.11}{\nano\m\per\femto\s}$. (b) Principle of SFBEI at an avoided crossing band structure. An electron starting in the valence band (blue circle) can undergo two different pathways within one optical cycle: a Landau-Zener transition after a quarter of an optical cycle followed by three-quarters of intraband motion, or vice versa. The final state occupation (red circle) depends on the interference of both pathways. (c) Electric field $\mathbf{E}(t)$ (red line) and its associated vector potential $\mathbf{A}(t)$ of a few-cycle light pulse. For this showcase, a 1.2~cycle waveform at \SI{800}{\nano\m} (1\,cycle $=\SI{2.7}{\femto\s}$) is employed with a peak optical field strength $E_0=\SI{3}{\volt\per\nano\m}$ and a carrier-envelope phase (CEP) $\varphi_\text{CE}=-\pi/2$. (d) Instantaneous band gap $\varepsilon^\text{cv}(t)$ experienced by the electron-hole pair indicated in (a). We show three cases for the hopping parameter $\gamma$ as in (a). Time integration over the green shaded area yields the dynamic phase $\phi^\text{cv}(t)$ (modulo $\hbar$), which is shown on the right axis, starting at $\phi^\text{cv}(-2\,\mathrm{cycles})=0$ . (e) Absolute value of the associated transient interband dipole coupling $V_\mathbf{k}^\text{cv}=\mathbf{E}(t)\cdot\mathbf{d}^\text{cv}(\mathbf{k})$ (blue line) induced by $\mathbf{E}(t)$ (red line). We note that $\mathbf{d}^\text{cv}$ is independent of $\gamma$. Regions of likely Landau-Zener tunneling events refer to the maxima in $\left|V_\mathbf{k}^\text{cv}\right|$, which are indicated as gray dashed lines through panels (d)--(f). (f) Conduction band population $\rho^\text{c}$ resulting from the analytical model (full lines) and the TDSE (dashed lines) with $\gamma$ as indicated in (a). Crucially, the residual $\rho^\text{c}$ (circles) differ significantly even for slight changes in $\gamma$ due to the interferometrically high sensitivity of the electron's dynamic phase on band modifications.}
            \label{Fig_1}
		\end{center}
	\end{figure*}

	A particularly intriguing situation emerges when strong light fields drive Bloch electrons across and between the bands, thereby realizing strong-field Bloch electron interferometry (SFBEI)~\cite{Heide2021b}. Here, the optical electric field directs electrons on a momentum trajectory where they accumulate a quantum phase as the imprint of the respective band shape. In addition, the electron wavefunction is split between adjacent bands and recombined on a sub-optical-cycle time scale, giving rise to self-referenced interference [see \fig{Fig_1}(b)], the key to access the electron quantum phase~\cite{Kelardeh2015,Higuchi2017,Chizhova2017}. The outcome, a measurable electric current, may be harnessed to retrieve the underlying band structure with interferometric precision and a sub-cycle time scale of the driving laser. Because this is much faster than lattice motion, we expect this technique to allow deep insights into band structure variations due to phononic driving. 
	
    Here, we examine the quantum dynamics based on the graphene band structure where the essential electron dynamics appear close to the vertices of the Brillouin zone, the K points, where the valence and lowest conduction band form a Dirac cone-like dispersion relation, see \fig{Fig_1}(a) and Refs.~\cite{Ishikawa2013,Kelardeh2015,Higuchi2017}. When an intense femtosecond laser pulse impinges, it transiently exerts momentum to electrons in the valence and conduction bands. Hence in the strong-field regime, the dynamics are no longer directed by the laser pulse envelope but by the optical carrier electric field $\mathbf{E}(t)$ [linearly $x$-polarized, see \fig{Fig_1}(c), dashed vs. full red line]. The resulting intraband trajectory [\fig{Fig_1}(a), blue double arrows] of any coherently driven electron can be described by Bloch's acceleration theorem~\cite{Bloch1929} $\mathbf{k}(t)=\mathbf{k}_0+\tfrac{e}{\hbar}\mathbf{A}(t)$ with $\mathbf{A}(t)=-\int_{-\infty}^{t}\mathbf{E}(t')\mathrm{d}t'$ the vector potential [\fig{Fig_1}(c), blue line] associated with $\mathbf{E}(t)$, and $\mathbf{k}_0$ the initial wave vector. 
	
	It is well known that such a driven electron may undergo a Landau-Zener transition from one band to another when it approaches an avoided crossing formed by two bands. More precisely, while the driven electron is moving in a certain band, a coherent electron-hole pair [\fig{Fig_1}(a), red and blue circles] may be created once the electron undergoes an impulsive Landau-Zener interband transition~\cite{Landau1932,Zener1932}, where it tunnels in between the valence ($\mathrm{v}$) and conduction ($\mathrm{c}$) band whenever a strong dipole coupling $V^\text{cv}_\mathbf{k}(t)=\mathbf{E}(t)\cdot\mathbf{d}^\text{cv}(\mathbf{k})$ (instantaneous Rabi frequency modulo $\hbar$) is given [\fig{Fig_1}(e), blue line and shaded vertical bands stripes]. The transition dipole matrix element
	\begin{equation}\label{eq:dipole}
		\mathbf{d}^\text{cv}(\mathbf{k})=e\braketmatrix{\psi^\text{c}_\mathbf{k}(\mathbf{r})}{i\frac{\partial}{\partial\mathbf{k}}}{\psi^\text{v}_\mathbf{k}(\mathbf{r})}
	\end{equation}
	characterizes the overlap of Bloch states $\psi_\mathbf{k}^{(\alpha)}(\mathbf{r})$ ($\alpha=\text{c,\,v}$). In the case of graphene, $\mathbf{d}^\text{cv}(\mathbf{k})$ strongly increases towards singular values at the K points, giving rise to substantial excitation probability in the Dirac cones.

Because the optical driving field is oscillatory in nature, even for few-cycle pulses, the electron is repeatedly driven back and forth in momentum space. This combined with the non-zero Landau-Zener transition probabilities close to the avoided crossing gives rise to Landau-Zener-Stückelberg-Majorana (LZSM) interferometry~\cite{Landau1932,Zener1932,Stueckelberg1932,Majorana1932,Ivakhnenko2023}, which we here dub as SFBEI in the context of a solid~\cite{Kelardeh2015}.  Here, the Landau-Zener transition events form the electron beamsplitters, whose properties can be approximated by the LZSM formula (Eq.~7 of Ref.~\cite{Ivakhnenko2023}). For any interferometer, the phase acquired in the beam-split state between two transition events at times $t_1$ and $t_2$ is important. On the adiabatic intraband trajectory of electrons it is given by the dynamic phase: 
	\begin{equation}\label{eq:PhiD}
		\phi_\mathbf{k}^\text{cv}(t)=\frac{1}{\hbar}\int_{t_1}^{t_2}\varepsilon_\mathbf{k}^\text{cv}(t)\mathrm{d}t,
	\end{equation}
	which is just the integral over its instantaneous energy separation $\varepsilon_\mathbf{k}^\text{cv}(t)=\varepsilon_\mathbf{k}^\text{c}(t)-\varepsilon_\mathbf{k}^\text{v}(t)$ [\fig{Fig_1}(c)].
	
	This dynamic phase accumulation determines the outcome of the sub-cycle SFBEI: In the first half of an optical cycle [for instance, between $t=-0.5$ and 0~cycles in Figs.~\ref{Fig_1}(d)--(f)], the electron wavefunction is split into two pathways: part of the electron undergoes a Landau-Zener transition [\fig{Fig_1}(b), pathway 1] and part of the electron undergoes purely intraband motion in the valence band [\fig{Fig_1}(b), pathway 2]. In the second half-cycle ($t=0\dots0.5$~cycles), when the electric field reverses, intra- and interband dynamics are interchanged between the two pathways such that the identical final conduction band state is reached [\fig{Fig_1}(b), red circles]. 

	Depending on the difference in dynamic phase accumulated on the two pathways, constructive interference, i.e., net conduction band excitation $\rho^\text{c}$, or destructive interference (no excitation) may result. This example holds for a single optical cycle. When using a few-cycle optical pulse for driving, the residual excitation probability depends on more than two but multiple passages through the avoided crossing. This leads to a strongly non-monotonic evolution of $\rho^\text{c}(t)$ [\fig{Fig_1}(f); the underlying model is outlined below]. Hence, the outcome of residual conduction band excitation $\rho^\text{c}(t\rightarrow\infty)$ [\fig{Fig_1}(f), green circles] is a result of the intricate interplay of inter- and intraband dynamics.
	
	We now study the influence of both the driving laser's waveform and the band shape on SFBEI. For demonstration purposes we introduce a first-order perturbative rate model derived from the time-dependent Schrödinger equation (TDSE) to provide an intuitive understanding of the relevant dynamics. We use a nearest-neighbor tight-binding model as described in~\cite{Saito1998}. 

		This framework has been proven to capture the relevant dynamics correctly compared to state-of-the-art time-dependent density functional theory~\cite{Li2021}. Moreover, by employing the overlap integral matrix described in Ref.~\cite{Saito1998}, the valence and conduction band are sculpted more precisely with their energetic symmetry lifted toward the $\Gamma$ point. 
		
		The TDSE
		\begin{equation}\label{eq:TDSE}
			i\hbar\frac{\mathrm{d}\Psi_\mathbf{k}(t)}{\mathrm{d}t}=\mathcal{H}_\mathbf{k}(t)\Psi_\mathbf{k}(t)
		\end{equation} 
		is solved using the wave function $\Psi_\mathbf{k}(t)=\left[a_\mathbf{k}^\text{v}(t),\,a_\mathbf{k}^\text{c}(t)\right]$ and the length gauge representation of the Hamiltonian
		\begin{equation}\label{eq:Ham-length}
			\mathcal{H}_\mathbf{k}(t)=\begin{bmatrix}
				0 & -V_\mathbf{k}^\text{cv}(t)e^{i\phi_\mathbf{k}^\text{cv}(t)} \\
				-V_\mathbf{k}^\text{cv}(t)e^{-i\phi_\mathbf{k}^\text{cv}(t)} & 0
			\end{bmatrix}.
		\end{equation}
		\Eq{eq:TDSE} can be solved numerically to yield an exact solution to $\rho_\mathbf{k}^\text{c}(t)=\left|a_\mathbf{k}^\text{c}(t)\right|^2$ [\fig{Fig_1}(f), dashed lines].
		
		To obtain a more intuitive understanding of the dynamics and the ensuing conduction band excitation, we rewrite \eq{eq:TDSE} as $\Psi_\mathbf{k}(t)=\mathcal{U}_\mathbf{k}(t)\Psi_\mathbf{k}(-\infty)$ with $\mathcal{U}_\mathbf{k}(t)=\mathcal{T}\exp\left[-\tfrac{i}{\hbar}\int_{-\infty}^{t}\mathcal{H}_\mathbf{k}(t')\mathrm{d}t'\right]$ and the time ordering operator $\mathcal{T}$. By expanding the above expression to first order as $\mathcal{U}_\mathbf{k}(t)\approx1-\tfrac{i}{\hbar}\int_{-\infty}^{t}\mathcal{H}_\mathbf{k}(t')\mathrm{d}t'$, the first-order perturbative excitation rate is obtained as~\cite{Bychkov1970,Krieger1986}
		\begin{equation}\label{eq:1storder}
			\rho_\mathbf{k}^\text{c}(t)=\left|-\frac{i}{\hbar}\int_{-\infty}^{t}V_\mathbf{k}^\text{cv}(t')\exp\left[i\phi_\mathbf{k}^\text{cv}(t')\right]\mathrm{d}t'\right|^2.
		\end{equation}
		Its temporal evolution is shown in \fig{Fig_1}(f) as full lines. It matches the results of the full TDSE solution well, which is important for the discussion to follow now. Clearly, the match can be improved further by  including higher orders in the form of a Dyson series~\cite{Kruchinin2018}, which is not needed here.
		
		This simplified rate reveals two key conditions for a net excitation into the conduction band to occur at an instant $t$: (1) A large dipole coupling $V_\mathbf{k}^\text{cv}(t)$, and (2) a constructive phase evolution. Condition (2) is fulfilled when a phase $\Delta\phi_\mathbf{k}^\text{cv}+\tilde{\phi}=2\pi n$ is accumulated between transition events [Figs.~\ref{Fig_1}(d)--(f), gray dashed lines] such that the complex-valued integrand of \eq{eq:1storder} does not vanish after time integration~\cite{Wismer2016,Higuchi2017,Shevchenko2010}. In fact, this requirement is equivalent to constructive LZSM interference~\cite{Ivakhnenko2023}. $\tilde{\phi}$ is a transition phase that can
		be approximated by the Stokes phase~\cite{Ivakhnenko2023,Kayanuma1997}; here it accounts to $\tilde{\phi}\approx\pi$. $n$ is an integer and can be linked to a multiphoton order (see detailed discussion below). Based on this model we now inspect the band shape sensitivity of the excitation process.
		
		The graphene tight-binding Hamiltonian depends on three numerical parameters only: The lattice constant $a$, the hopping parameter $\gamma$, and the overlap integral $s$ (see Ref.~\cite{Saito1998} for a definition; we use $s=0.129$ as a correct representation of the bands toward the $\Gamma$ point~\cite{Saito1998}. While the lattice constant $a=\SI{2.46}{\angstrom}$ is well known from first-principle computations~\cite{Gui2008}, the value of the hopping parameter $\gamma$ is less well determined, typically ranging from $\gamma=\SI{-3.03}{\eV}$ to \SI{-2.50}{\eV}~\cite{Saito1998,Reich2002,Gui2008}. We will show that we can measure $\gamma$ based on fitting our experimental data stemming from excitation close to the K points. 
		
		The inset of \fig{Fig_1}(a) illustrates the influence of $\gamma$ on the conduction band (valence band equivalently):  The Fermi velocity $v_\text{F}=\tfrac{3}{2}\left|\gamma\right|a$, i.e., the slope of the Dirac cone, is slightly increased when we tune $\gamma$ from \SI{-2.8}{\eV} to \SI{-3.0}{\eV}. This in turn is reflected in a minor change of instantaneous electron energy for a given $\mathbf{k}$-value [\fig{Fig_1}(a, d)]. Importantly, the resultant dynamic phase evolution clearly encodes this small variation of instantaneous energies within a few femtoseconds [\fig{Fig_1}(d), wrapped lines]. While the dipole transition matrix element is independent of $\gamma$, the resulting excitation band probability $\rho^\text{c}$ after the laser pulse has gone is completely changed, but can be fully understood in terms of the differing dynamic phase evolution [\fig{Fig_1}(e), green circles].
		
\begin{figure*}
			\begin{center}
				\includegraphics[width=\linewidth]{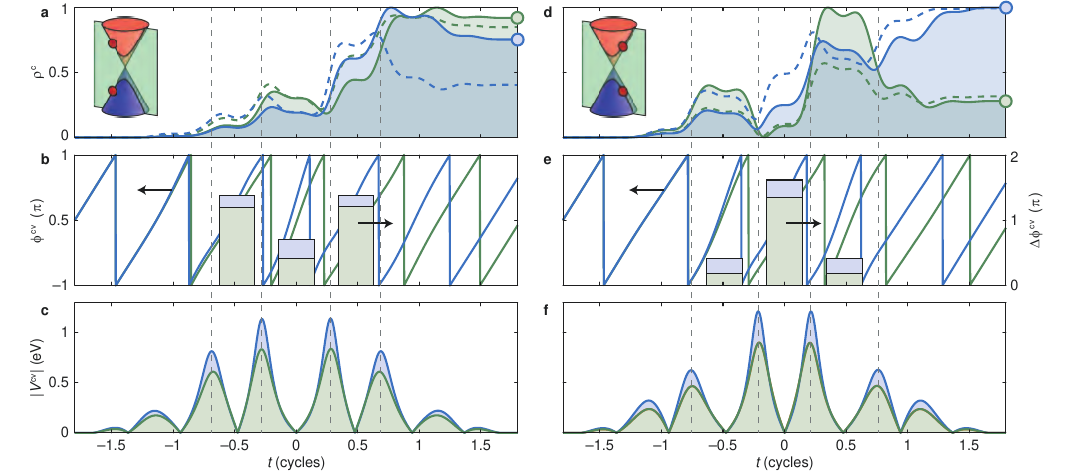}
				\caption{Field strength-dependent momentum symmetry breaking. (a) Conduction band population $\rho^\text{c}(t)$ obtained from the analytical model (full lines) and the TDSE (dashed lines) for an electron starting at $\mathbf{k}_0=[14.1,\,6.6]\,\si{\per\nano\m}$ where the next K point is located at $[14.7,\,8.5]\,\si{\per\nano\m}$ ($k_{0,x}<\mathrm{K}_x$, see inset). The identical pulse as shown in \fig{Fig_1}d is used with $\varphi_\text{CE}=\pi/2$ and $E_0=\SI{3}{\volt\per\nano\m}$ (green lines in all panels) and $E_0=\SI{4}{\volt\per\nano\m}$ (blue). (b) Dynamic phase $\phi^\text{cv}(t)$ (full lines) and the phase difference $\Delta\phi^\text{cv}$ (bars) accumulated in between instants of interband tunneling, indicated by the gray dashed lines. (c) Absolute value of the interband dipole coupling $V^\text{cv}$. The gray dashed lines are positioned at the maxima of $\left|V^\text{cv}\right|$. (d)--(f) Identical description of the dynamics shown in (a)--(c) but for an electron starting at $\mathbf{k}_0=[15.3,\,6.6]\,\si{\per\nano\m}$ ($k_{0,x}>\mathrm{K}_x$ with identical distance to the K point, see inset in (d)). The imbalance of the residual $\rho^\text{c}$ between $k_{0,x}<\mathrm{K}_x$ and $k_{0,x}>\mathrm{K}_x$ [circles in (a) and (d)] reverses going from $E_0=\SI{3}{\volt\per\nano\m}$ to $E_0=\SI{4}{\volt\per\nano\m}$, showing that the field strength-dependent dynamic phase evolution determines the direction of current flow.}
				\label{Fig_3}
			\end{center}
		\end{figure*}

		Furthermore, by controlling the temporal symmetry of the optical waveform by means of its carrier-envelope phase (CEP) $\varphi_\text{CE}$ we lift the population inversion symmetry of the Dirac cones. For example, by choosing $\varphi_\text{CE}=\pi/2$ the driving field forces electrons starting at two different initial wave vectors $\mathbf{k}_0$ with $k_{0,x}<\mathrm{K}_x$ and $k_{0,x}>\mathrm{K}_x$ on trajectories that lead to a different dynamic phase evolution and sequence of interband transition events (see Ref.~\cite{Heide2019} and discussion below around \fig{Fig_3}). As a consequence, the asymmetric momentum distribution $\rho^\text{c}$ gives rise to a ballistic current density
		\begin{equation}\label{eq:current}
			j=2e\sum_{\alpha}\int_{\text{BZ}}\hbar^{-1}\frac{\partial\varepsilon^{(\alpha)}(\mathbf{k})}{\partial k_x}\rho_\mathbf{k}^{(\alpha)}(t\rightarrow\infty)\frac{\mathrm{d}\mathbf{k}}{(2\pi)^2},
		\end{equation}
		which can be precisely controlled in amplitude and direction depending on the value of $\varphi_\text{CE}$~\cite{Higuchi2017,Heide2020,Boolakee2022}. The factor 2 in \eq{eq:current} accounts for both electron spins; the integration is performed across the complete Brillouin zone (BZ).
		
\begin{figure}
			\begin{center}
				\includegraphics[width=89mm]{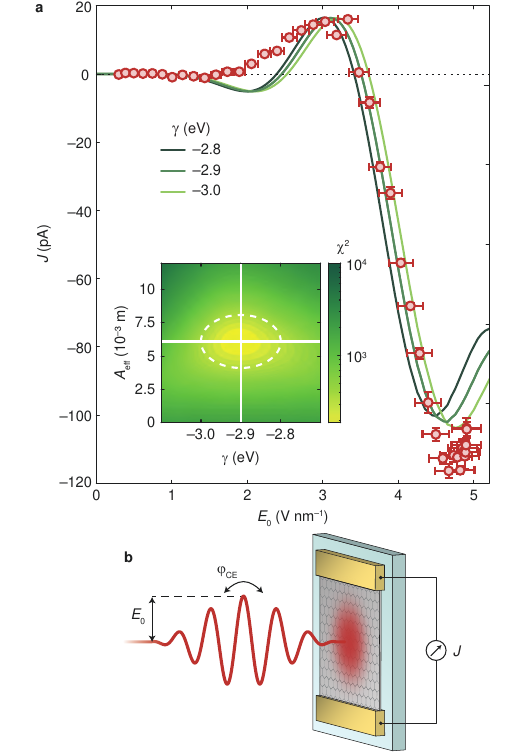}
				\caption{Carrier-envelope phase-dependent current in monolayer graphene. (a) Experimentally obtained CEP-dependent current (red circles with error bars) as a function of the optical peak field strength $E_0$. Error bars indicate the standard deviation of the mean of 33 independent samples. The three green lines are obtained from numerical TDSE computations with the absolute current magnitude as only free parameter, with hopping parameters $\gamma$ as indicated. Inset, $\chi^2$ distribution from fitting the simulated data to the experimentally obtained data points. The dashed white ellipse indicates the $1\sigma$ interval around the global minimum marked by the full white lines. (b) Sketch of the experimental setup: Epitaxially grown monolayer graphene on a SiC substrate is illuminated with tightly focused CEP-stable 2-cycle laser pulses. Induced CEP-dependent currents are measured via two gold electrodes attached to the graphene.}
				\label{Fig_2}
			\end{center}
		\end{figure}

		To demonstrate the feasibility of band structure retrieval based on experimentally obtained field-driven current, we excited epitaxial monolayer graphene on silicon carbide with \SI{5}{\femto\s} (full width at half maximum, equal to 1.9 optical cycles) CEP-stable laser pulses centered at \SI{800}{\nano\m} (photon energy of \SI{1.55}{\eV}). We controlled the result of SFBEI, i.e., the residual distribution of $\rho_\mathbf{k}^\text{c}$, by modulating the CEP of the pulse train at a carrier-envelope offset frequency $f_\text{CEO}=\SI{3.3}{\kilo\hertz}$ and by varying the peak optical field strength $E_0$ from 0 to \SI{5}{\volt\per\nano\m} on the substrate surface. The pulses are focused to a $1/e^2$ intensity radius of \SI{1.8}{\micro\m} in the center of a $5\times\SI{2}{\micro\m\squared}$ graphene strip. The graphene strip is attached to two gold electrodes to measure the CEP-sensitive photo-induced currents [\fig{Fig_2}(b)]. Field-induced currents are isolated from the CEP-insensitive background via dual-phase lock-in detection referenced to $f_\text{CEO}$ following transimpedance current amplification with \SI{e7}{\volt\per\ampere}. See Ref.~\cite{Boolakee2022} for experimental and sample fabrication details.
		
		\Fig{Fig_2}(a) shows the measured current (red circles) as a function of $E_0$. Whereas in a previous experiment we were able to show this current for up to \SI{3}{\volt\per\nano\m}~\cite{Higuchi2017}, here we can extend the data to \SI{5}{\volt\per\nano\m} (\SI{0.5}{\volt\per\angstrom}) to reveal the peculiar oscillatory evolution of the current. In addition to the previously observed current reversal below \SI{2}{\volt\per\nano\m} (visible here at around \SI{1.5}{\volt\per \nano\m} from negative to positive current), which had been identified as an onset marker of the strong-field regime~\cite{Higuchi2017,Heide2021b}, we report a second current reversal at around \SI{3.5}{\volt\per\nano\m} followed by a steep growth to a maximum absolute value of \SI{116(2)}{\pico\ampere} at \SI{4.8(2)}{\volt\per\nano\m}. Towards \SI{5}{\volt\per\nano\m}, the current magnitude shows the onset of a decline towards zero again.
		
		The characteristic current scaling is well reproduced by the TDSE simulation [\eq{eq:TDSE}]. We show three results with slightly different values of $\gamma$ [\fig{Fig_2}(a), green lines]. All required numerical parameters are taken from the experiment. The simulated current was obtained by averaging over arbitrary lattice angles with respect to the laser polarization to account for all possible orientations of epitaxial grains within the laser focus~\cite{Emtsev2009}. Furthermore, focal averaging is applied to include the spatial Gaussian field distribution in the focal plane. Most importantly, the two sign changes as well as the relative current amplitude between the maxima at 3 and \SI{4.8}{\volt\per\nano\m} are well reproduced. An increase in $\left|\gamma\right|$, i.e., an increase of the Fermi velocity [\fig{Fig_1}(a), inset] mainly results in a shift of features to higher $E_0$. 
		
        The qualitative and quantitative agreement together with the clearly visible $\gamma$ dependence allows us to determine the optimal hopping parameter with a fitting procedure. For this, we sweep $\gamma$ and adjust the effective magnitude $A_\text{eff}$ to match $J_\text{TDSE}=A_\text{eff}\,j$ with the experimentally observed current magnitude $J$. The inset of \Fig{Fig_2}(a) shows the $\chi^2$ distribution
		\begin{equation}\label{eq:chi}
			\chi^2(\gamma,A_\text{eff})=\sum_{n}\frac{\left[J^{(n)}-J_\text{TDSE}^{(n)}(\gamma,A_\text{eff})\right]^2}{\left|J^{(n)}\right|},
		\end{equation}
		with $n$ indicating all measured points with $E_0>\SI{1}{\volt\per\nano\m}$, yielding $A_\text{eff}=\SI{6(2)e-3}{\m}$ and $\gamma=\SI{-2.9(1)}{\eV}$ as the best fit result. The corresponding Fermi velocity equals $v_\text{F}=\SI{1.07(4)}{\nano\m\per\femto\s}$. We note that $A_\text{eff}$ incorporates the signal transmission to the electrodes and is thus a highly setup-dependent parameter.
  
        This Fermi velocity fits remarkably well with previously obtained values obtained by infrared spectroscopy [\SI{1.02(1)}{\nano\m\per\femto\s}]~\cite{Orlita2008}, scanning tunneling microscopy [\SI{1.070(6)}{\nano\m\per\femto\s}] \cite{Miller2009}, and ARPES [\SI{1.00(5)}{\nano\m\per\femto\s}]~\cite{Sprinkle2009}. We note that all these values are obtained from epitaxial graphene on SiC, exactly like in our case. Hence, these results prove that strong-field Bloch electron interferometry is a viable approach for band structure retrieval.
		
		It is intriguing to realize that the experimental value for the Fermi velocity results from integrating the dynamical phase. Because the integral result matches independently obtained Fermi velocity results, it appears appropriate to look into the quantum dynamics as it unfolds. \Fig{Fig_3} illustrates the showcase of transient electron dynamics driven by a 1.2-cycle \SI{800}{\nano\m} pulse with $\varphi_\text{CE}=\pi/2$ (as for \fig{Fig_1}) based on the analytic rate [\eq{eq:1storder}]. Here we compare the dynamics for electrons starting left [(a)--(c)] and right [(d)--(f)] of the K point [see insets of Figs.~\ref{Fig_3}(a) and (d)]. The occurrence of symmetry breaking for $\varphi_\text{CE}=\pi/2$ becomes immediately apparent by the different temporal spacings between interband transition events [maxima in $\left|V^\text{cv}(t)\right|$, panels (c) vs. (f)]. Moreover, the residual momentum imbalance reverses by increasing the optical field strength from $E_0=3$ to $\SI{4}{\volt\per\nano\m}$ [Figs.~\ref{Fig_3}(a) and (d), green and blue circles] -- similar to the experimentally observed second current reversal shown in \fig{Fig_2}(a).
		
		This behavior can be largely explained by the dynamic phase accumulation $\Delta\phi^\text{cv}$ [Figs.~\ref{Fig_3}(b) and (e), bar graphs] between subsequent transition events as previously outlined in the discussion of \eq{eq:1storder}. Here $\Delta\phi^\text{cv}$ is shown between the four main transition events (gray dashed lines). As an example, for $E_0=\SI{3}{\volt\per\nano\m}$ the electron at $k_x>\text{K}_x$ experiences a sequence $\Delta\phi^\text{cv}$ that can be roughly estimated as $[0,\,\pi,\,0]$ [\fig{Fig_3}(e), green bars]. Together with $\tilde{\phi}\approx\pi$, a succession of destructive, constructive, and destructive interference manifests in the course of $\rho^\text{c}(t)$ at $t\approx[-0.25,\,+0.25,\,+0.75]$~cycles [\fig{Fig_3}(d), green line]. 
		
		By increasing $E_0$ to \SI{4}{\volt\per\nano\m}, $\Delta\phi^\text{cv}$ is increased by approximately $\pi/4$ in the respective time windows [\fig{Fig_3}(e), blue bars], leading to a substantial change in $\rho^\text{c}(t)$ [\fig{Fig_3}(d), blue line]. This dynamic phase evolution is more detuned from the optimum constructive or destructive resonance conditions: going from $t=-0.5$\,cycles to $t=0$, $\rho^\text{c}$ barely changes, followed by a moderate increase toward $t=+0.5$\,cycles despite a significant increase in $\left|V^\text{cv}\right|$ [\fig{Fig_3}(f), blue line]. Going to $t=+1$\,cycle, again rather little change is obtained. After passage of the pulse, the conduction band state is fully occupied while for $E_0=\SI{3}{\volt\per\nano\m}$ only \SI{28}{\percent} filling is achieved. Again, the full TDSE model supports this analytic analysis [\fig{Fig_3}(d), dashed lines]. The same analysis can be applied to the complementary point at $k_x<\text{K}_x$ to unveil its $E_0$-dependent excitation probability, and finally the emergence of the $E_0$-dependent momentum asymmetry.
		
		To compute the resulting current density $j$ at the respective optical field strengths, the dynamics of all initial wave vectors across the Brillouin zone must be taken into account. Hence, the complex interplay of final states competing for the net momentum imbalance determine the resultant value of $j$~\cite{Heide2020}. In addition, with the longer pulses employed in the experiment, the interference of even more interband transition events contributes to the composition of final outcome, which also leads to more complex inter-cycle dynamics. Importantly, the sensitivity of the quantum phase to the band shape is preserved, and the interference method continues to be well applicable for band structure retrieval as demonstrated in our experiment.
		
		We expect the full potential of SFBEI-based band structure retrieval to unfold (1) when the full quantum phase space spanned by the dynamic and the geometric phase is probed, and (2) when dynamic changes to the band structure on the femtosecond time scale arise. In example, the Berry curvature of gapped 2D materials gives rise to a non-trivial geometric phase for electrons encircling the K point that may be probed subsequently as a field-driven Hall current that enables probing the underlying band structure~\cite{Mitra2023}. Similarly, the topology landscape of Weyl semi-metals, Moiré and patterned dielectric superlattices~\cite{Andrei2021, Forsythe2018}, and topological insulators, as well as correlations between Bloch electrons~\cite{Freudenstein2022}, could be probed based on strong-field electron interferometry~\cite{Ma2017,Nematollahi2019}. As for time-resolved band retrieval, transient deformations of the lattice, such as coherent optical phonons~\cite{Neufeld2022}, may become visible from the induced currents with a time resolution given by the optical cycle duration of the probing laser pulse. 

		To summarize, we extend the generation of field-driven currents in monolayer graphene up to an unprecedented peak optical field strength of $E_0=\SI{5}{\volt\per\nano\m}$, thereby revealing an oscillating evolution of the current as a function of $E_0$. We can clearly show that this current results from strong-field Bloch electron interferometry, based on electrons undergoing complex coherent intraband motion coupled with interband transitions. An analytical model helps us to understand the underlying quantum dynamics, whereas we utilize the observed current dependence matched by TDSE computation results to retrieve the graphene Fermi velocity as \SI{1.07(4)}{\nano\m\per\femto\s}, in excellent agreement with previously experimentally obtained values. We expect this method to measure with femtosecond temporal precision band structures with high accuracy in a plethora of recently emerging quantum materials.
		
		\vspace{2ex} This work has been supported in part by the Deutsche Forschungsgemeinschaft (SFB 953 “Synthetic Carbon Allotropes”), the PETACom project financed by Future and Emerging Technologies Open H2020 program, ERC Grants NearFieldAtto and AccelOnChip.

	\end{document}